\documentclass{article} 

\usepackage{graphicx}
\graphicspath{{converted_graphics/}}
\begin{document}

\begin{center}
{\LARGE A Convection Chamber for Measuring}\vskip6pt

{\LARGE Ice Crystal Growth Dynamics}\vskip16pt

{\Large Kenneth G. Libbrecht}\footnote{%
e-mail address: kgl@caltech.edu; see also
\par
http://www.its.caltech.edu/\symbol{126}atomic/publist/kglpub.htm}{\Large \
and Helen C. Morrison}\vskip4pt

{\large Department of Physics, California Institute of Technology}\vskip-1pt

{\large Pasadena, California 91125}\vskip-1pt

\vskip18pt

\hrule \vskip1pt \hrule
\vskip 14pt
\end{center}

\noindent \textbf{Abstract. We present the design of a general-purpose
convection chamber that produces a stable environment for studying the
growth of ice crystals from water vapor in the presence of a background gas.
Crystals grow in free fall inside the chamber, where the temperature and
supersaturation are well characterized and surprisingly uniform. As crystals
fall and land on a substrate, their dimensions are measured using direct
imaging and broad-band interferometry. We also present a parameterized model
of the supersaturation inside the chamber that is based on differential
hygrometer measurements. Using this chamber, we are able to observe the
growth and morphology of ice crystals over a broad range of conditions, as a
function of temperature, supersaturation, gas constituents, gas pressure,
growth time, and other parameters.}

\section{\noindent The Convection Chamber}

The formation of complex structures during solidification often results from
a subtle interplay of nonequilibrium, nonlinear processes, for which
seemingly small changes in molecular dynamics at the nanoscale can produce
large morphological changes at all scales. One popular example of this
phenomenon is the formation of snow crystals, which are ice crystals that
grow from water vapor in an inert background gas. Although this is a
relatively simple physical system, snow crystals display a remarkable
variety of columnar and plate-like forms, and much of the phenomenology of
their growth remains poorly understood, even at a qualitative level \cite%
{libbrechtreview}.

To perform quantitative experimental investigations of ice crystal formation
and morphological instabilities, one must create a stable growth
environment, typically an inert background gas with a fixed temperature and
humidity. For studying growth in near-atmospheric conditions, this requires
temperatures of typically -40 C to 0 C, pressures of order 1 bar, and
supersaturations of typically 0-20 percent. Since a nonzero supersaturation
is an intrinsically nonequilibrium state, producing well-determined growth
conditions with sufficient accuracy and stability is often a significant
experimental challenge.

We have constructed a basic convection chamber for this purpose, shown
schematically in Figure 1. The chamber consists of a stainless steel vacuum
tank with an inside diameter of 53 cm and an inside height of 81 cm, and
thus an interior volume of 179 liters. The outside of the chamber is covered
(including the sides and both ends) by 3-mm thick copper plates with
soldered cooling pipes. A programmable chiller cools the plates by
circulating methanol through the soldered pipes. Once the system is stable,
the chiller can maintain the interior temperature of the chamber down to -35
C with an accuracy of about 0.1 C.

\begin{figure}[tbp] 
  \centering
  \includegraphics[width=4.5in,height=3.95in,keepaspectratio]{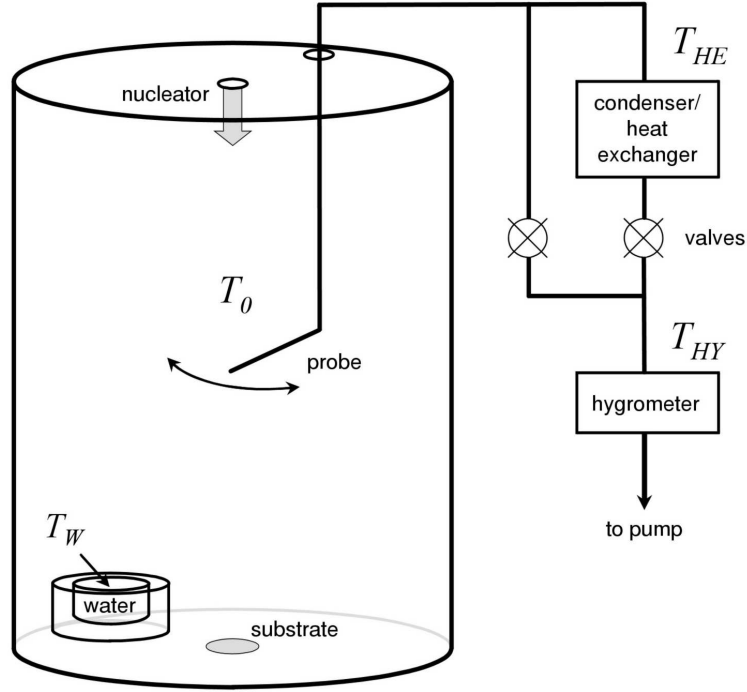}
  \caption{Schematic diagram of the growth chamber and differential hygrometer, as
described in the text. The probe and water reservoir inside the chamber are
drawn roughly to scale, whereas the hygrometer elements have been enlarged
for clarity.}
  \label{fig:Tank}
\end{figure}

An insulated water reservoir, also shown in Figure 1, is placed at the
bottom of the tank to provide a source of water vapor. The water reservoir
has an inside diameter of 14.5 cm and a depth of 5 cm, and it is typically
filled with 0.8 liters of deionized water. The water temperature is servo
controlled to an accuracy of approximately 0.25 C. The heated water drives
convection inside the chamber that mixes water vapor into the surrounding
gas. In steady state this simple configuration provides a stable environment
in which the temperature and supersaturation are surprisingly uniform over
much of the interior of the chamber.

A pulse of rapidly expanding gas is used to nucleate the growth of ice
crystals at the top of the chamber \cite{hallett}. The nucleator is made
from a 5-cm-long pipe, 2 cm in diameter, with solenoid-actuated valves on
both ends, connected to a source of compressed gas that has been saturated
with water vapor. The pipe and valves are kept at approximately the same
temperature as the chamber via simple conduction to the exterior cold
plates. The first valve is opened to admit compressed gas into the pipe and
then closed. The second valve is then opened to discharge the compressed gas
into the growth chamber. The rapid expansion cools the saturated gas inside
the pipe to nucleate small ice crystals \cite{hallett}. We typically use
nitrogen gas in the nucleator, although argon, helium, and other gases work
as well. Typically about 10 psi is needed, with higher gas pressures
producing more crystals.

The crystals grow in free fall in the supersaturated environment until a
combination of gravity and convective currents causes some of them to fall
onto a glass substrate at the bottom of the chamber. At this point the
crystal dimensions can be measured from below using direct imaging and
interferometry \cite{precision}, described further below. This versatile
chamber can produce a large number of crystals in a short time, allowing one
to build up statistics on the growth and morphology as a function of
temperature, supersaturation, gas constituents, gas pressure, growth time,
and other parameters.

The temperature and supersaturation inside the chamber are measured using
the probe and differential hygrometer set-up shown in Figure 1. The
temperature probe is a calibrated thermistor with an absolute accuracy of
0.1C, located at the end of a sample tube placed halfway between the top and
bottom of the chamber. Under typical conditions (chamber temperature $T_{0}$
between 0 C and -25 C with a pressure of one bar), we have found that the
temperature is uniform to approximately 0.1 C as the probe is moved from
side to side, even though on one side of the probe's travel it is directly
over the water reservoir. The walls of the chamber are typically 2 C cooler
than $T_{0}$ (since the warmer water adds heat to the gas, which must be
removed via contact with the walls), but this is only evident from the probe
when it is within 2 cm of the walls. Convection efficiently mixes the gas
inside the chamber to maintain this nearly uniform temperature.

Supersaturation is measured using a Vaisala HUMICAP hygrometer consisting of
a thin-film polymer material that absorbs water vapor depending on the
surrounding humidity, which is then determined by electronically measuring
the dielectric properties of the film. The hygrometer sensor is placed
inside a temperature-regulated enclosure kept at temperature $T_{HY}$ (see
Figure 1), and gas entering this enclosure has its temperature brought to $%
T_{HY}$ before it contacts the hygrometer sensor.

A pump draws gas from the probe tube inside the chamber to the hygrometer at
a rate of approximately 3 cm$^{3}$/sec, either directly or by first passing
it through a condenser/heat-exchanger, as shown in Figure 1. (In operation,
one valve is open while the other is closed.) The condenser/heat-exchanger
consists of a thin copper tube soldered to a heat sink at temperature $%
T_{HE}.$ The water content of gas that has passed through the
condenser/heat-exchanger is equal to that of saturated gas at temperature $%
T_{HE}.$ Care is taken that water does not condense anywhere inside the
differential hygrometer except in the condenser/heat-exchanger. In
particular, we heat the inside of the thin teflon probe tube inside the
chamber using a coaxial heating wire, and we make sure all other flow tubes
are at temperatures $T>T_{0}.$

We model the operation of this differential hygrometer by assuming the
hygrometer measures%
\[
H=\frac{c_{input}}{c_{sat}(T_{HY})}+A 
\]%
where $c_{input}$ is the water content of the gas entering the hygrometer, $%
c_{sat}(T_{HY})$ is the water content of saturated gas at the hygrometer
temperature $T_{HY},$ and $A$ represents some nearly constant offset in the
instrument. For a perfect hygrometer, we would have $A=0,$ so the formula
then gives simply the relative humidity.

If we send gas from the probe directly into the hygrometer, we measure 
\[
H_{1}=\frac{c_{probe}}{c_{sat}(T_{HY1})}+A 
\]%
where $c_{probe}$ is the water content of the gas entering the probe. If the
gas first goes through the heat exchanger, then we measure 
\[
H_{2}=\frac{c_{HE}}{c_{sat}(T_{HY2})}+A 
\]%
where $c_{HE}=c_{sat}(T_{HE})$ is the water content after the gas has gone
through the heat exchanger. Putting these together, we have%
\[
c_{probe}=\left[ \left( H_{1}-H_{2}\right) c_{sat}(T_{HY2})+c_{sat}(T_{HE})%
\right] \frac{c_{sat}(T_{HY1})}{c_{sat}(T_{HY2})} 
\]%
and this in turn gives the supersaturation inside the tank%
\[
\sigma =\frac{c_{probe}-c_{sat}(T_{probe})}{c_{sat}(T_{probe})} 
\]%
By using a differential measurement, we eliminate $A$ and thus increase the
accuracy of our measurement.

In operation we first set $T_{HE}\approx T_{0}$ and set $T_{HY}$ so that the
water content of the gas in the chamber will produce a relative humidity
(after being heated to $T_{HY}$) of approximately 50 percent. We then flow
gas from the probe directly to the hygrometer and monitor $H_{1},T_{HY1},$
and $T_{0}$ until these quantities have stabilized (the hygrometer
relaxation time is typically several minutes at temperatures near -15 C).
Next we switch the valves and monitor $H_{2},T_{HY2},$ and $T_{HE}$ until
the readings are again stable, and then we repeat this sequence several
times. A single hygrometer measurement sequence may take approximately one
hour to complete, and our results are shown in Figure 2.

\begin{figure}[tb] 
  \centering
  \includegraphics[width=4.5in,height=3.64in,keepaspectratio]{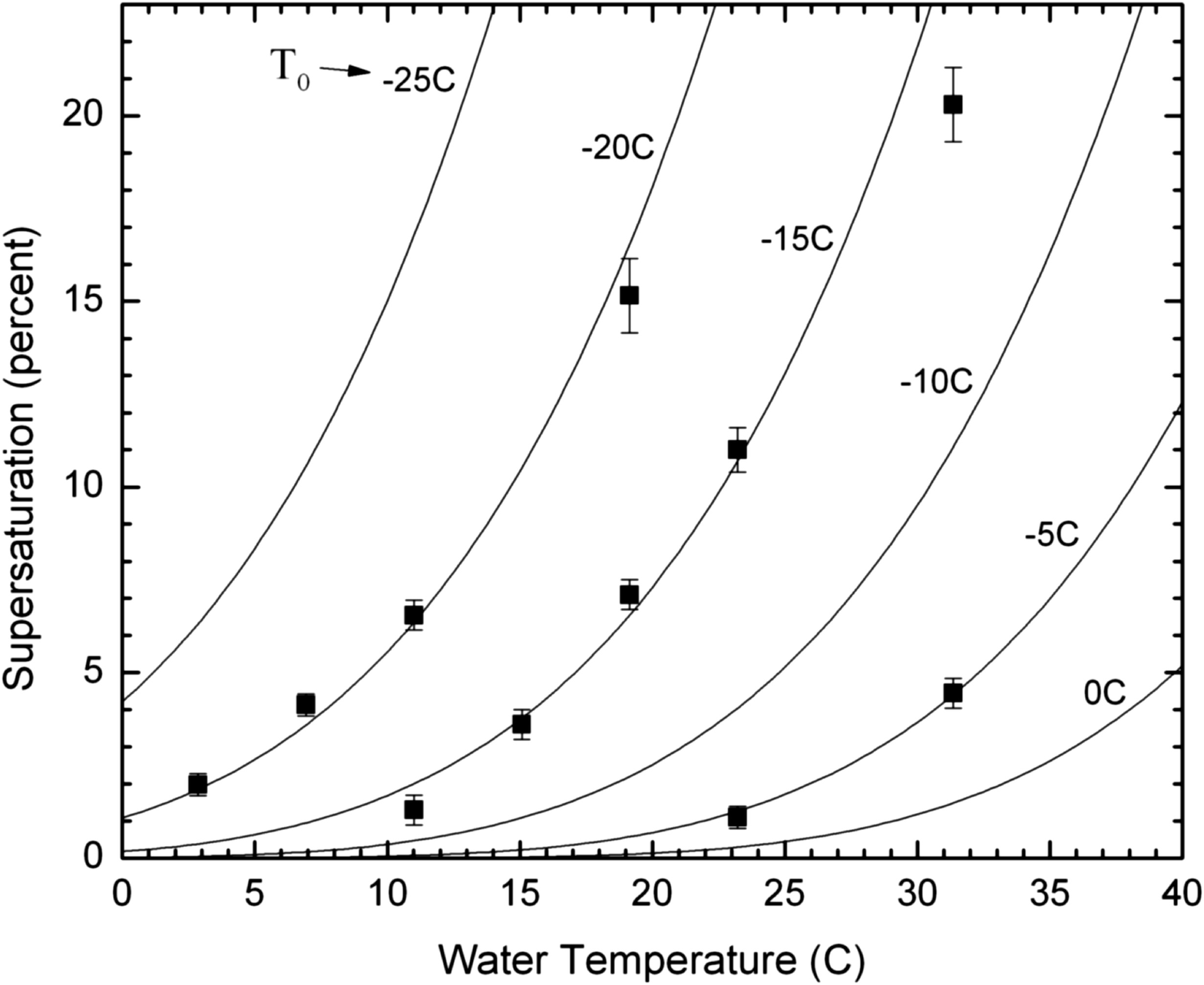}
  \caption{Supersaturation measurements taken in air at a
pressure of one bar, at chamber temperatures of $T_{0}=$ -5 C, -15 C, and
-20 C (points), along with model calculations at several temperatures
(lines).}
  \label{fig:HNew}
\end{figure}

\section{Modeling Supersaturation}

We model the rate of evaporation of water into a closed chamber by \cite%
{hirth, zemansky} 
\[
\frac{dc}{dt}\propto g(T_{W},T_{0})[c_{sat}(T_{W})-c_{sat}(T_{0})] 
\]%
where the temperatures are defined above and in Figure 1. Here we include a
dimensionless function $g(T_{W},T_{0})$ to describe the strength of
convection inside the chamber. When $\Delta T=T_{W}-T_{0}$ is small, the
resulting convection is weak, so little water vapor is carried into the
chamber, whereas more vigorous convection will increase $dc/dt$. Thus we
expect $g(T_{W},T_{0})$ will depend mainly on $\Delta T$ and will increase
monotonically with $\Delta T.$

In addition to evaporation, we also have condensation onto the walls, which
we describe as%
\[
\frac{dc}{dt}\propto \sigma c_{sat}(T_{0})
\]%
In steady-state, these two rates will be equal, giving%
\[
\sigma =g(T_{W},T_{0})\frac{c_{sat}(T_{W})-c_{sat}(T_{0})}{c_{sat}(T_{0})}
\]%
To fit our measurements, we use%
\[
g(T_{W},T_{0})=2.6(1-e^{-\Delta T/30})^{\alpha }
\]%
with $\alpha =(0.11T_{0}+5.5),$ where $\Delta T$ and $T_{0}$ are measured in
degrees Celsius. Although this functional form for $g(T_{W},T_{0})$ is 
\textit{ad hoc }and includes several adjustable parameters, the resulting
model fits our data well and provides physically reasonable trends with $%
T_{W}$ and $T_{0}.$ We expect that the detailed form of $g(T_{W},T_{0})$
would likely change with a different chamber geometry or over a different
temperature range. Using this model, we are able to interpolate between our
hygrometer measurements to determine the supersaturation inside the growth
chamber as a function of $T_{0}$ and $T_{W}$ to an absolute accuracy of
about 20 percent.

\begin{figure}[tbp] 
  \centering
  \includegraphics[width=4.5in,height=3in,keepaspectratio]{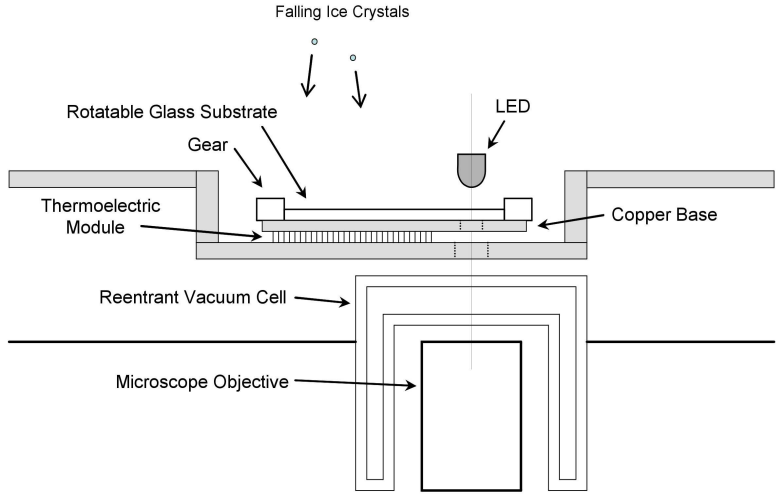}
  \caption{The substrate
assembly at the bottom of the convection chamber. Ice crystals grow in free
fall inside the chamber and some will fall onto the rotatable glass
substrate where they can be measured.}
  \label{fig:optics1}
\end{figure}

\section{Measuring Ice Crystals}

After nucleation, ice crystals grow in free-fall inside the chamber until a
combination of air currents and gravity carries some of them to a
temperature-controlled glass substrate at the bottom of the chamber (see
Figure 1). A schematic view of the substrate assembly is shown in Figure 3.
The substrate itself is an uncoated glass window that has been glued to the
center of a 5-cm-diameter Delrin spur gear. The substrate lies flat on a
temperature-regulated copper base, and a drive gear (not shown in Figure 3)
rotates the substrate about its vertical axis. An LED with a small diffusing
screen provides illumination.

The rotating substrate solves a number of problems we had encountered
earlier using a fixed substrate. First, the LED provides ample illumination
from a large solid angle without impeding the fall of crystals to most of
the substrate. Calculations show that heating from the LED is negligible.
Second, the observable annulus on the substrate has a large area compared to
the microscope's field of view, making it possible to find crystals to
observe even when their surface density is low. This is important because
nucleating a large number of crystals can significantly reduce the
supersaturation inside the chamber, adding uncertainty to the growth
measurements. Finally, this geometry makes it relatively easy to clean the
substrate. This is accomplished by inserting a moistened Q-tip on a long rod
from the top of the chamber. A mechanical guide (not shown in Figure 3)
positions the Q-tip at one point on the observable annulus of the substrate.
Once the Q-tip is resting on the glass, we rotate the substrate to drag-wipe
the surface. 

We use a Mitutoyo M Plan Apo 10X microscope objective, since this provides
roughly 1-$\mu $m resolution with a 30-mm working distance. The objective is
mounted on a 3-axis stage, and two video cameras provide low- and
high-magnification views, the former with a field of view of approximately 3
mm. The low-magnification view is useful for finding crystals on the
substrate, which are then observed more closely at high magnification. A
reentrant evacuated cell with coated windows provides insulation between the
microscope objective, which is at room temperature, and the cold substrate.
A second copper base beneath the substrate base plate is separately cooled
to ensure that the temperature of all parts surrounding the substrate remain
uniform and stable during observations.

\begin{figure}[tbp] 
  \centering
  \includegraphics[width=4.5in,height=2.49in,keepaspectratio]{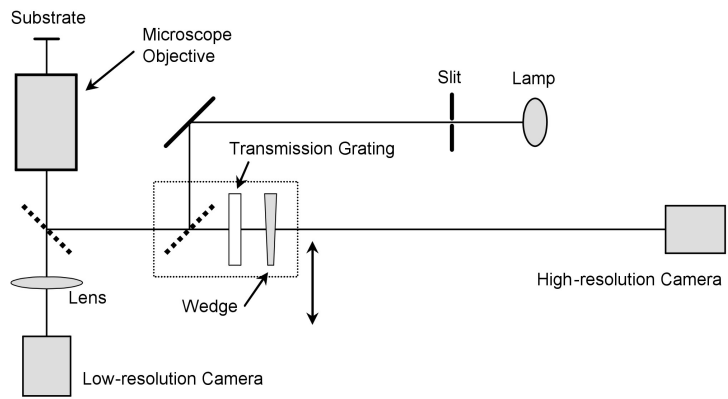}
  \caption{Optical layout for measuring
the size and thickness of crystals landing on the substrate. For direct
imaging at high and low resolution, the transmission grating assembly (shown
in a box) is moved out of the optical path. When this assembly is in place
(shown), the high-resolution camera records the broad-band interferometer
spectrum described in the text.}
  \label{fig:optics2}
\end{figure}

Thin, plate-like ice crystals tend to land flat on the substrate, with their
broad faces contacting the glass, so their thickness cannot be easily
measured using microscopy from below. Furthermore, the plates often have
thicknesses of 1-2 $\mu $m, which is too small to be accurately measured
using direct imaging. For such crystals, we use broad-band optical
interferometry to measure thickness, with the optical layout shown in Figure
4. Light from an incandescent bulb passes through a rectangular slit, and
the slit is imaged by the microscope objective onto the ice crystal. Some of
the incident light reflects off the glass-ice boundary, and more reflects
off the ice-air boundary. Amplitudes of the two reflections add with an
additional phase equal to $\varphi =4\pi hn/\lambda ,$where $h$ is the ice
thickness and $n(\lambda )$ is the index of refraction of ice at wavelength $%
\lambda $. The resulting reflected intensity is%
\begin{eqnarray*}
I &\sim &\left\vert R_{1}+R_{2}e^{i\varphi }\right\vert ^{2} \\
&\sim &(1+r^{2})+2r\cos \varphi \\
&\sim &(1+\cos \frac{4\pi hn}{\lambda })
\end{eqnarray*}%
where $r=R_{2}/R_{1}$ is the ratio of reflection amplitudes, which depends
on the index jumps at the two interfaces. We have taken $r\approx 1$ for the
last expression because the index of refraction of ice is approximately
midway between those of air and glass. Note that the limit $h\rightarrow 0$
gives the maximum reflected intensity, because the air-glass interface has
the largest index jump and thus produces the largest reflected intensity.

\begin{figure}[tbp] 
  \centering
  \includegraphics[width=4.5in,height=4.56in,keepaspectratio]{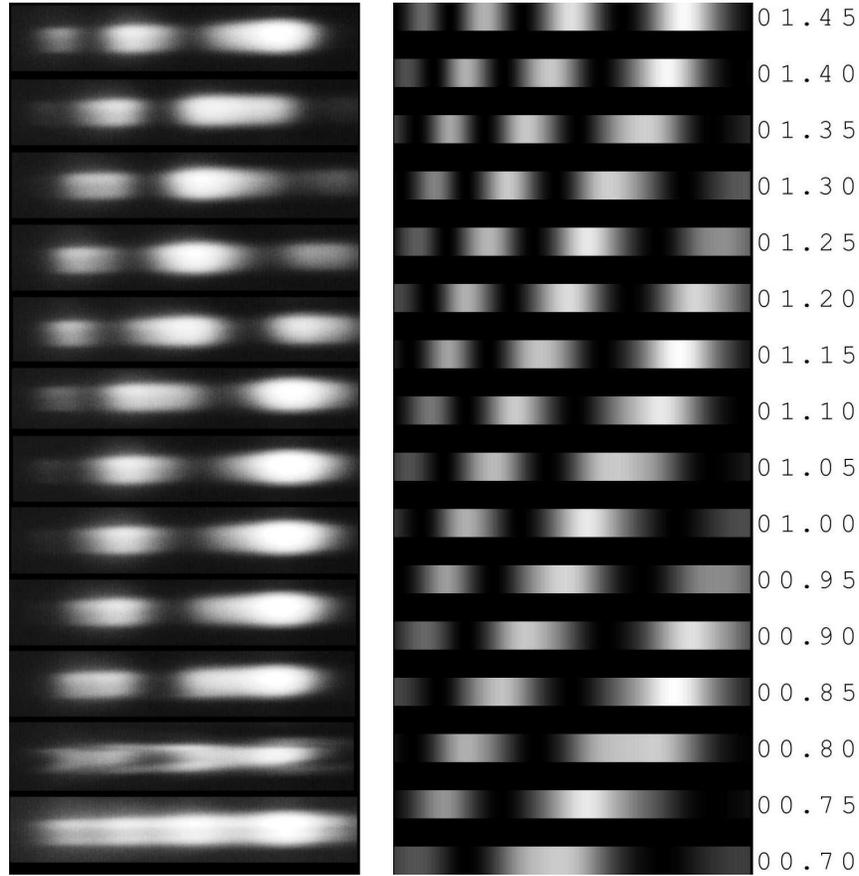}
  \caption{The left panel shows a series
of interferometer spectra taken as a plate-like ice crystal slowly
evaporated from the substrate. The right panel shows a series of calculated
spectra, each labeled with the modeled ice thickness in microns. In both
cases the spectra cover the range from 440 to 820 nm. Comparing the top
spectrum on the left with the set of calculated spectra reveals that the
crystal started with a thickness of approximately 1.1 $\protect\mu $m. After
some evaporation, the 10th spectrum shows the crystal with a thickness of
0.8 $\protect\mu $m. After this, the lateral size of the crystal diminished
to where it was smaller than the imaged slit, giving the next-to-last
spectrum. The last spectrum was taken after the crystal had evaporated away
completely. Here the residual brightness variations result from the lamp
spectrum along with weak accidental etalons within the optics.}
  \label{fig:ifo}
\end{figure}

The reflected light is dispersed by a transmission grating (see Figure 4),
and the interference spectrum is viewed by the high-resolution camera.
Figure 5 shows a series of interference spectra taken as a plate-like
crystal slowly evaporated away, along with a set of calculated spectra.
Calibration for the calculations was obtained by shining red ($\lambda =654$
nm) and green ($\lambda =532$ nm) laser pointers through the slit and
observing where the resulting spots appeared on the video image.

We use three techniques to infer crystal thickness from the interferometer
spectra: 1) direct comparison of the video image with model spectra, as
shown in Figure 5; 2) simply counting fringes on the video image; and 3)
measuring the fringe spacing at the red end of the spectrum on the video
image. When the crystal thickness is $h<2$ $\mu $m, direct comparison with
model spectra is the only technique that yields accurate data, but this
method is quite slow to use. When the thickness is $1<h<4$ $\mu $m, simply
counting fringes on the video image yields data with roughly 10 percent
accuracy, which is sufficient for our requirements. However, counting cannot
be used for still thicker crystals because the fringes tend to merge
together at the green end of the spectrum. Therefore, when $3<h<15$ $\mu $m,
we measure the fringe spacing at the red end of the spectrum on the video
image. With proper calibration of the three techniques, we are able to
quickly and easily measure crystal thicknesses over the range $0.5<h<15$ $%
\mu $m with overall absolute accuracies of 20 percent or better.

\section{Operation}

A typical run with this apparatus begins with sealing the chamber and
cooling it to the set temperature, which takes several hours. The air may be
pumped out and replaced with another gas as desired. When the temperature is
stable, we add water to the reservoir and wait for it to reach its set
temperature. In general, the temperature of the gas inside the chamber
depends on both the temperature of the chamber walls (the chiller set point)
and the water temperature.

Care must be taken to ensure that the chamber walls are covered fairly
uniformly with ice, in order to produce well-defined boundary conditions for
realizing a stable and known supersaturation. When operating the chamber
above -15 C, we first cool the chamber to -15 C for several hours to make
sure any water droplets condensed on the walls are frozen. Also, we
repeatedly nucleate many crystals immediately after adding room-temperature
water to the reservoir, so that some will attach to the chamber walls.

The number of crystals produced by a single cycle of the nucleator is a
strong function of the nucleator gas pressure, with higher pressures
producing more crystals. We typically run with 5-15 psi and adjust this so
it is within 1-2 psi of the point when crystals first appear. By measuring
the mass of a typical falling crystal and the number of crystals on the
substrate after a single nucleation cycle, we estimate that the
supersaturation is not substantially reduced by the growing crystals,
provided we keep the nucleation pressure low.

Once an ice crystal lands on the substrate, it will grow or evaporate with
time, depending on the substrate temperature. In practice, we have found
that the substrate temperature must remain within a few hundredths of a
degree of the zero-growth temperature to keep crystals from growing
significantly after 1-2 minutes on the substrate. We minimize the adverse
effects of growth or evaporation on the substrate by observing crystals as
quickly as possible after they have fallen. After nucleating a set of
crystals inside the chamber, we monitor the substrate while searching for
fallen crystals, continually recording the video output to DVD. Each time a
crystal is spotted on the low-res camera, it is moved to the high-res camera
for measurement, and its thickness is recorded with the interferometer,
which all takes just a few seconds. The growth time is assumed to be the
elapsed time between nucleation and observation, ignoring the short
residence time of crystals on the substrate. When crystals first begin to
fall, we measure them all; after about a minute we record only the larger
ones for a short additional time. This strategy usually allows us to record
1-2 dozen crystals in a single nucleation cycle. If observations of long
growth times are desired, we use the same measurement strategy, except we
use a shutter to keep crystals off the substrate for some period of time
before beginning observations. Under typical conditions, all the crystals
will fall after 5-10 minutes.

Our experience with this general-purpose convection chamber indicates that
it can produce a stable environment in which the temperature and
supersaturation are easily controlled and surprisingly uniform. Ice crystals
grown in the chamber exhibit a good uniformity with respect to size and
morphology. Measurement accuracies of 10-20\% (in crystal size and
thickness, as well as chamber supersaturation) are routinely achieved. Since
the chamber is easy to use and can produce a wide variety of environmental
conditions, we are finding it quite useful for investigating the detailed
physics of ice crystal growth.

\section{Acknowledgements}

We acknowledge support for HCM by the Summer Undergraduate Research
Fellowship (SURF) program and the CamSURF program at Caltech.

\end{document}